\documentclass[aps,prd,reprint,nofootinbib,amsmath,amssymb]{revtex4-2}
\usepackage[a4paper,margin=0.75in]{geometry}
\usepackage{graphicx}
\usepackage{dcolumn}
\usepackage{bm}
\usepackage{hyperref}
\usepackage{xcolor}

\usepackage{makecell}
\usepackage{hyperref}

\usepackage{booktabs}   
\usepackage{multirow}   

\begin{document}


\title{Neural-Network Identification of LIGO Noise Transients Using Binary Auxiliary-Channel Activity}

\author{D. Chintala${}^{1}$}
\author{J. Temple${}^{1}$}
\author{S. Shapiro${}^{1}$}
\author{L. Wilson${}^{1}$}
\author{K. Wayt${}^{1}$}
\author{M. Van Keuren${}^{1}$}
\author{P. Neubauer${}^{1}$}
\author{K. Rose${}^{1}$}
\author{D. Moffa${}^{1}$}
\author{L. Wade${}^{1}$}
\author{M. Wade${}^{1}$}

\affiliation{${}^1$Department of Physics, Kenyon College, 201 N College Rd, Gambier, OH 43022, USA}


\begin{abstract}
Transient noise artifacts (``glitches") in the Laser Interferometer Gravitational-Wave Observatory (LIGO), Virgo and KAGRA (LVK) detectors can mimic astrophysical signals and reduce confidence in gravitational-wave detections. 
The LVK's iDQ pipeline identifies these glitches by analyzing correlations between auxiliary sensors and the strain channel using the Ordered Veto List (OVL) algorithm. 
We revisit neural-network approaches to this problem using modern training infrastructure and introduce two architectures: GIANTS, a single network trained on all auxiliary-channel features, and TITAN, a mixture-of-experts model composed of subsystem-specific networks and a learned combiner. 
Evaluated on ten independent test periods spanning LIGO's third observing run, both models outperform OVL on most test sets, with TITAN providing the most consistent performance over time. 
We further show that nearly all predictive information contained in Omicron trigger features is captured by a simple binary indicator of auxiliary-channel activity. 
Networks trained on these binary inputs achieve essentially the same performance as models using the full 8,440-dimensional feature set while reducing the input dimensionality by an order of magnitude. 
These results demonstrate that modern neural networks provide a robust alternative to OVL and suggest that substantially simpler data representations are sufficient for accurate glitch identification in future iDQ deployments.
\end{abstract}

\maketitle

\section{Introduction}
The Laser Interferometer Gravitational-Wave Observatory (LIGO) consists  of two US-based interferometers with unparalleled levels of sensitivity used to detect gravitational waves that manifest as nearly imperceptible changes in detector strain \cite{LIGOScientific:2014pky}.
To date, these detectors, along with the Virgo and KAGRA detectors \cite{AdvVirgoPaper, kagradet2}, have successfully identified well over 300 gravitational-wave events with $p_{\rm astro} > 0.5$ from merging compact objects \cite{GWTC-5}.
Due to LIGO’s high sensitivity, transient noise artifacts (``glitches'') caused by terrestrial activity burden the gravitational-wave search process by mimicking real signals and lowering confidence in detections.
LIGO detectors employ auxiliary sensors at each site to monitor experimental and environmental factors that may impact the strain data quality \cite{LIGO:2021ppb, LIGO:2024kkz, LIGO:2026ika}.

The immense amount of time series data produced by the auxiliary sensors can be reduced using time-frequency decomposition methods.
One such method, employed by the Omicron \cite{Robinet_2020} trigger generator, is a Q-transform \cite{Qtransform1, Qtransform2}.  When applied to the auxiliary sensor data channels, the time series data can be reduced to a set of features that describe the activity in these channels over time.
The LVK has employed a system known as iDQ to predict the presence of a glitch in gravitational-wave strain data based on patterns in auxiliary channel activity at each interferometer \cite{Essick_2021, 2024arXiv241204638H}.
Data quality information produced by iDQ has been successfully incorporated into the ranking statistics for modeled searches for gravitational-wave signals \cite{2020arXiv201015282G, PhysRevD.106.102006}.
Ref.~\cite{Biswas_2013} explored the use of a few different machine learning algorithms (MLAs), including an Artificial Neural Network (ANN), for making these predictions and found that the Ordered Veto List (OVL) algorithm was as accurate as, and less computationally expensive than, the other MLAs explored.
For this reason, OVL is the primary algorithm used by iDQ.

In this work, we revisit the use of ANNs for predicting the presence of glitches in the main gravitational-wave strain channel using auxiliary sensor data with more modern ANN software infrastructure that smoothly incorporates hyperparameter optimization.
We explore two unique MLAs, which we refer to as GIANTS and TITAN, for use in the iDQ pipeline and compare their performances to each other and to OVL.
GIANTS is a single neural network trained on the auxiliary-channel features from all detector subsystems at once.
TITAN is built from twelve neural networks, one per detector subsystem, so that each network specializes in the glitch couplings characteristic of that subsystem.
Each network sees only the auxiliary channels belonging to its subsystem. A final ``combiner" network then takes the twelve subsystem outputs and produces the overall glitch classification.
We find that, despite their different architectures, TITAN and GIANTS achieve comparable performance on a held-out test set drawn from the same period as the training data. 
Because interferometer noise is non-stationary \cite{LIGOScientific:2019hgc}, however, we further evaluate both models on ten independent stretches of data from different times in the observing run, and find that TITAN retains its performance more consistently than GIANTS. 
We also compare directly to OVL, using the area under the ROC curve (AUC) of glitch-detection efficiency versus false-alarm probability. 
Our best-performing model, a variant of TITAN trained on the binary inputs described below, achieves a higher AUC than OVL on nine of the ten independent test sets, with OVL performing better on the remaining one.

In addition, we revisit a feature-importance study \cite{Biswas_2013} to determine which Omicron trigger features are most informative for predicting glitches in the gravitational-wave strain channel. 
We find that the detailed feature values carry little predictive information. 
Our MLAs perform comparably when given only a binary indicator of whether each auxiliary channel showed activity, defined by Omicron's signal-to-noise-ratio threshold, and no information about that activity, such as the values of the ten Omicron features.
This indicates that a much simpler, binary representation of auxiliary-channel activity is sufficient for glitch identification at no cost in performance.

We discuss the inputs and outputs, training and evaluation, and methods for choosing hyperparameters for the ANNs used in this analysis in Sec.~\ref{sec:ann}.
In Sec.~\ref{sec:models}, we describe the architectures of our two models and report their baseline performance on a held-out test set.
We present a brief feature study in Sec.~\ref{sec:features} and show the effect of providing just a single binary feature to our MLAs and the two new models that come from this method. 
We then compare each model in Sec.~\ref{sec:ovl_comp} to each other and the OVL classifier.
We conclude by interpreting our results in Sec.~\ref{sec:discussion} and discuss future directions.


\section{Neural Network Methodology}
\label{sec:ann}
 In this work, we use several ANNs for binary classification of glitches in LIGO data. This section describes the inputs and outputs of these ANNs, their training and evaluation, and the methods used to choose their hyperparameters. We also clarify some vocabulary used throughout this paper.

\subsection{Data Representation}

The inputs for the ANNs in this analysis are values of trigger features calculated by Omicron from various safe auxiliary channels at a given integer second time.
Auxiliary channels monitor the LIGO instruments and their environments.
A ``safe'' auxiliary channel is one whose activity cannot be affected by the presence of a gravitational wave, which is why they can be safely used to predict the presence of glitches.
The ANN inputs correspond either to a time when there is a glitch in the main gravitational-wave strain channel or to a time when there is not a glitch in the main gravitational-wave strain channel, which is called a ``clean'' time.
A glitch in this context refers to transient noise in the main gravitational-wave strain channel that exceeds some signal-to-noise ratio (SNR) threshold and is within some specified frequency band of interest.
Trigger features are recorded during times when the activity in an auxiliary channel exceeds some SNR threshold in one or more time-frequency tiles.
If an auxiliary channel's activity does not exceed this threshold during the 1-second window of a glitch or clean event within the frequency band of interest, then the ten features in that auxiliary channel are all set to nominal default values, which are chosen to be zero in most cases.
The ten features calculated by Omicron are outlined in Tab.~\ref{tab:ligo_features} and details of the Omicron algorithm on how these features are computed can be found in Ref.~\cite{Robinet_2020}.

\begin{table}[h!]
\centering
\begin{tabular}{|c|l|}
\hline
\textbf{Name} & \textbf{Trigger Features} \\
\hline
amplitude & Estimated amplitude \\
\hline
SNR & Estimated amplitude SNR \\
\hline
tstart & Starting time of tile \\
\hline
tend & Ending time of tile \\
\hline
fstart & Starting frequency of tile  \\
\hline
fend & Ending frequency of tile \\
\hline
time & Central time of tile \\
\hline
frequency & Central frequency of tile \\
\hline
phase & Phase value of tile \\
\hline
$Q$ & The $Q$ plane $Q$ value \\
\hline
\end{tabular}
\caption{The list of Omicron trigger features for a time-frequency tile.}
\label{tab:ligo_features}
\end{table}

For the networks presented here, we pulled data from 844 safe auxiliary channels from the LIGO Hanford detector.
Of Omicron's ten features (Tab.~\ref{tab:ligo_features}), we replace the raw time feature with the iDQ-derived feature \texttt{deltatime}, which is the difference between the central time of the Omicron tile and the central time of the glitch (or clean) window. 
The default value for \texttt{deltatime}, applied when an auxiliary channel's activity does not exceed the SNR threshold during the 1-second window of interest, was chosen to be $-3$ rather than zero, because zero would indicate direct coincidence between the tile and window times. 
Each ANN input therefore comprises ten features per channel, Omicron's nine features plus \texttt{deltatime}, and an ANN using all 844 safe channels has $844 \times 10 = 8,440$ inputs. 
An ANN that uses fewer than 844 channels has proportionally fewer inputs.

The safe auxiliary channels used in our study are organized into twelve distinct subsystems relating to the properties of the LIGO detector.
Tab.~\ref{tab:ligo_subsystems} presents a complete list of the LIGO Hanford subsystems used in this study.
Some examples of auxiliary channels that are part of the PEM subsystem are those associated with accelerometers and magnetometers, which serve distinct but complementary roles in monitoring environmental noise.
Accelerometer channels, such as \texttt{H1:PEM-CS\_ACC\_BSC1\_ITMY\_X}, measure vibrations near key optics like the input test mass and help identify how ground motion or mechanical disturbances could couple into the main strain signal.
Magnetometer channels, such as \texttt{H1:PEM-CS\_MAG\_EBAY\_LSCRACK\_X}, track magnetic field fluctuations around electronics racks, where power supplies or external sources may introduce noise into actuators or electronics.
Both types of sensor feed into coupling-function analyses, which estimate how strongly environmental disturbances affect the strain, and their data are crucial for diagnosing glitches and applying vetoes \cite{abbott2016detchar, PEM_ref}.

\begin{table}[h!]
\centering
\begin{tabular}{|c|l|}
\hline
\textbf{Code} & \textbf{Subsystem Name} \\
\hline
ASC & Alignment Sensing and Control \\
\hline
HPI & Hydraulic External Pre-Isolator \\
\hline
ISI & Internal Seismic Isolation \\
\hline
PEM & Physical and Environmental Monitoring \\
\hline
SUS & Suspension \\
\hline
PSL & Pre-Stabilized Laser \\
\hline
TCS & Thermal Compensation System \\
\hline
SQZ & Squeezing Subsystem \\
\hline
LSC & Length Sensing and Control \\
\hline
OMC & Output Mode Cleaner \\
\hline
IMC & Input Mode Cleaner \\
\hline
CAL & Calibration Subsystem \\
\hline
\end{tabular}
\caption{The list of safe LIGO Hanford auxiliary channel subsystems used in this study for trigger features.}
\label{tab:ligo_subsystems}
\end{table}

Each ANN works as a binary classifier with a single output between zero and one.
A classification threshold is applied to the output.
If a value is greater than the threshold, then the event is classified as a glitch; if the value is less than the threshold, then the event is classified as clean.  

\subsection{Training and Evaluation}
\label{sec:training}

In order to train and evaluate the performance of our ANNs, we used data from LIGO's third observing run (O3) \cite{O3_gwosc}.
Since Omicron had already been run over all auxiliary channels during the entirety of O3, we built a dataset by collecting all Omicron triggers above an SNR of 6.0 associated with glitches in the main strain channel during four disjoint, roughly week-long intervals of O3 (7.2 to 8.6 days each; Tab.~\ref{tab:clustering-counts-train}), which we refer to as Weeks 1–4.
We added to this set an equal number of randomly selected clean events so that our full dataset was balanced to include 50\% cleans and 50\% glitches.
Our dataset consisted of 27,182 events in total, each with 8,440 total auxiliary channel features.
All ANNs in this work were implemented in Python using the TensorFlow machine learning framework with the Keras API~\cite{tensorflow2015-whitepaper,chollet2015keras}.
We set aside 90\% of this dataset for training and validation and reserved 10\% for model evaluation or ``testing".

\begin{table}[h!]
\centering
\begin{tabular}{|l|c|c|r|}
\hline
\textbf{} & \textbf{Start (UTC)} & \textbf{End (UTC)} & \textbf{Glitches} \\
\hline
Week 1 & 2019-04-01 & 2019-04-08 & 2{,}293 \\
\hline
Week 2 & 2019-08-24 & 2019-09-01 & 1{,}447 \\
\hline
Week 3 & 2019-11-10 & 2019-11-18 & 8{,}859 \\
\hline
Week 4 & 2020-02-23 & 2020-03-02 & 992 \\
\hline
\end{tabular}
\caption{Per-week UTC date ranges and glitch counts for the four disjoint O3 training intervals. Dates give the UTC day on which each interval begins and ends; the intervals do not start or end at midnight and span 7.2, 7.8, 8.6, and 7.9 days, respectively. Training sets are 50/50 balanced by construction.}
\label{tab:clustering-counts-train}
\end{table}

Within the 90\% of the dataset used for the training process, we mark 10\% of the data (2,446 events) for validation and use the remaining data (22,018 events) directly for model training.
The validation set is used to determine how long we train our ANNs.
After every epoch of training, we evaluate the ANN on the validation set by computing its validation loss, which is a numerical measure of the distance between the model's predictions and the correct labels.
Training concludes when the validation loss stops noticeably improving or begins to get worse due to overfitting of our training set.
The model from the epoch with the lowest validation loss is saved.


The testing dataset (2,718 events) is used to quantify the saved model's efficacy.
The efficacy of a given ANN is quantified by the area under the ROC curve (AUC), which plots glitch-detection efficiency versus false-alarm probability.
Efficiency (true positive rate) refers to the fraction of glitches correctly labeled as glitches, and the false-alarm probability (false positive rate) is the fraction of cleans falsely labeled as glitches.
Each point on the ROC curve is a measurement of efficiency and false-alarm probability for a different glitch classification threshold.
The AUC is the integral of the ROC curve; the higher the AUC, the better the classifier.
Since efficiency and false-alarm probability each range from zero to one, a perfect classifier will have an ${\rm AUC}=1$, while a completely random classifier would measure ${\rm AUC}\approx0.5$.

To test the robustness of our ANNs to changes in the detectors over time, we built ten additional test sets of O3 Omicron triggers at LIGO Hanford, each spanning three to four days.
For each test set, we pulled all of the glitches from that time and an equal number of cleans from the same time to balance the set.
These ten test sets are outlined in Tab.~\ref{tab: test_list}.

      
\begin{table}[h!]
\centering
\begin{tabular}{|l|c|c|r|}
\hline
 & \textbf{Start (UTC)} & \textbf{End (UTC)} & \textbf{Glitches} \\
\hline
Test set 01 & 2019-07-14 & 2019-07-16 & 615 \\
\hline
Test set 02 & 2019-12-08 & 2019-12-10 & 1{,}186 \\
\hline
Test set 03 & 2020-01-18 & 2020-01-20 & 974 \\
\hline
Test set 04 & 2019-05-10 & 2019-05-12 & 550 \\
\hline
Test set 05 & 2019-06-15 & 2019-06-17 & 1{,}354 \\
\hline
Test set 06 & 2019-07-20 & 2019-07-23 & 549 \\
\hline
Test set 07 & 2019-09-08 & 2019-09-10 & 942 \\
\hline
Test set 08 & 2019-12-13 & 2019-12-15 & 7{,}598 \\
\hline
Test set 09 & 2020-01-10 & 2020-01-12 & 7{,}782 \\
\hline
Test set 10 & 2020-02-05 & 2020-02-07 & 382 \\
\hline
\end{tabular}
\caption{Per-set UTC date ranges and glitch counts for the O3 test intervals. Each test set is 50/50 balanced. Start and end dates are inclusive, so each set spans three full UTC days, except test set 06, which spans four.}
\label{tab: test_list}
\end{table}

\subsection{Hyperparameter Tuning}
\label{sec:hyper-tuning}

ANNs consist of model parameters and hyperparameters.
Model parameters are chosen by the algorithm during training, whereas hyperparameters are chosen in advance by the user.
ANN weights and biases are an example of model parameters.
They are initially randomized and then tuned during training in order to improve a model's efficacy.
The activation function between layers, on the other hand, is a hyperparameter chosen before training begins.
The performance of an ANN can be dramatically improved by selecting appropriate hyperparameters for the problem.
Though effective hyperparameters can be chosen manually, improvements are often achieved by employing a hyperparameter tuning algorithm.
In this work, we used the Keras Bayesian Optimization Tuner to hyperparameter tune our ANNs~\cite{dewancker2016bayesianoptimizationmachinelearning, chollet2015keras}.
Tab.~\ref{tab:params} provides a list of hyperparameters and hyperparameter options over which we tuned our ANNs.
Our tuning simulations ran for 100 trials with 1,000 epochs per trial with the objective to minimize the loss of the validation data.
We found that our models tended to generalize better over our ten test sets when optimizing on validation loss as opposed to validation accuracy.

\begin{table}[h!]
\centering
\renewcommand{\arraystretch}{1.2}
\begin{tabular}{|c|l|}
\hline
\textbf{Hyperparameters} & \textbf{Options} \\
\hline
Layers & 1 -- 6 \\
\hline
Nodes & 32 -- 512 \\
\hline
Dropout & 0.0 -- 0.8 \\
\hline
Activation & \makecell[l]{``elu'', ``gelu'', ``linear'',\\ ``relu'', ``selu'', ``softplus'',\\ ``softsign'', ``swish'', ``tanh''} \\
\hline
Learning Rate & 0.01, 0.001, 0.0001 \\
\hline
Optimizer & \makecell[l]{``SGD'', ``RMSprop'', ``Adam'',\\ ``Adadelta'', ``Adagrad'',\\ ``Adamax'', ``Nadam''} \\
\hline
Regularizer & ``L1'', ``L2'', ``L1\_L2'' \\
\hline
Regularizer Weight & 0.01, 0.001, 0.0001 \\
\hline
\end{tabular}
\vspace{0.5em} 
\caption{Hyperparameters that were tuned and their options and ranges.}
\label{tab:params}
\end{table}

\section{Model Architectures}
\label{sec:models}
Section~\ref{sec:ann} described the inputs, training and evaluation, and hyperparameter tuning procedures shared by every network in this study. 
Here we present the two models built on that foundation. 
GIANTS uses a single neural network that sees the auxiliary-channel features from all detector subsystems at once.
TITAN instead distributes the problem across twelve subsystem-specific networks, one per detector subsystem, whose outputs are merged by a final combiner network. 
These two designs let us test whether subsystem specialization improves performance beyond what a single hyperparameter-tuned ANN can achieve.
We close the section by comparing the baseline performance of both models on the four-week O3 dataset of Sec.~\ref{sec:training}.

\begin{figure}[h!]
    \centering
    \includegraphics[width=1.0\linewidth]{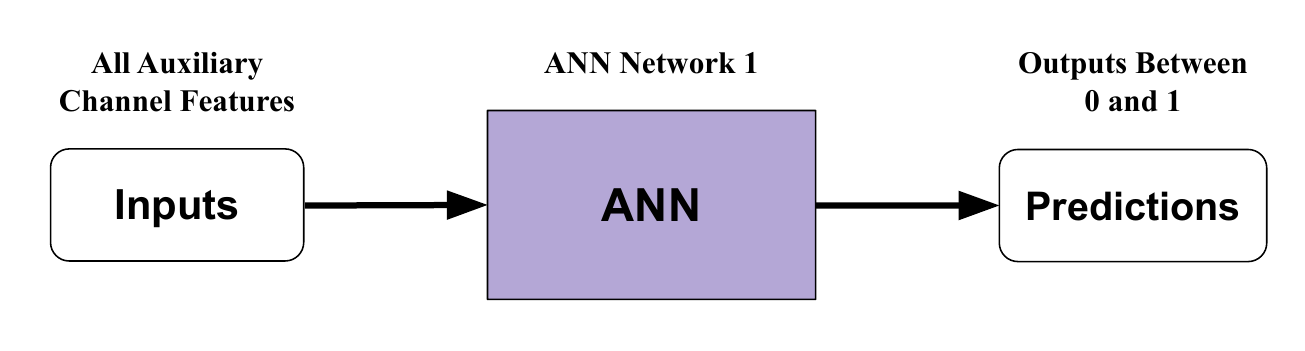}
    \caption{Flowchart of the GIANTS workflow. All auxiliary-channel features are passed to a single ANN, which outputs a glitch-classification score between 0 and 1.}
    \label{fig:giantsflowchart}
\end{figure}

GIANTS (Glitch Identification Algorithm using Neural-neTworkS) is a single neural network built to test whether advances in hyperparameter-tuning algorithms improve ANN performance on this problem. It receives 8,440 inputs (ten features for each of the 844 safe auxiliary channels) and returns a single glitch-classification score between 0 and 1. Its tuned hyperparameters are listed in Tab.~\ref{tab:giants_vs_binary_giants} of App.~\ref{sec:appendix_b}, and its workflow is illustrated in Fig.~\ref{fig:giantsflowchart}.

\begin{figure*}[t!]
    \centering
    \includegraphics[width=1.0\linewidth]{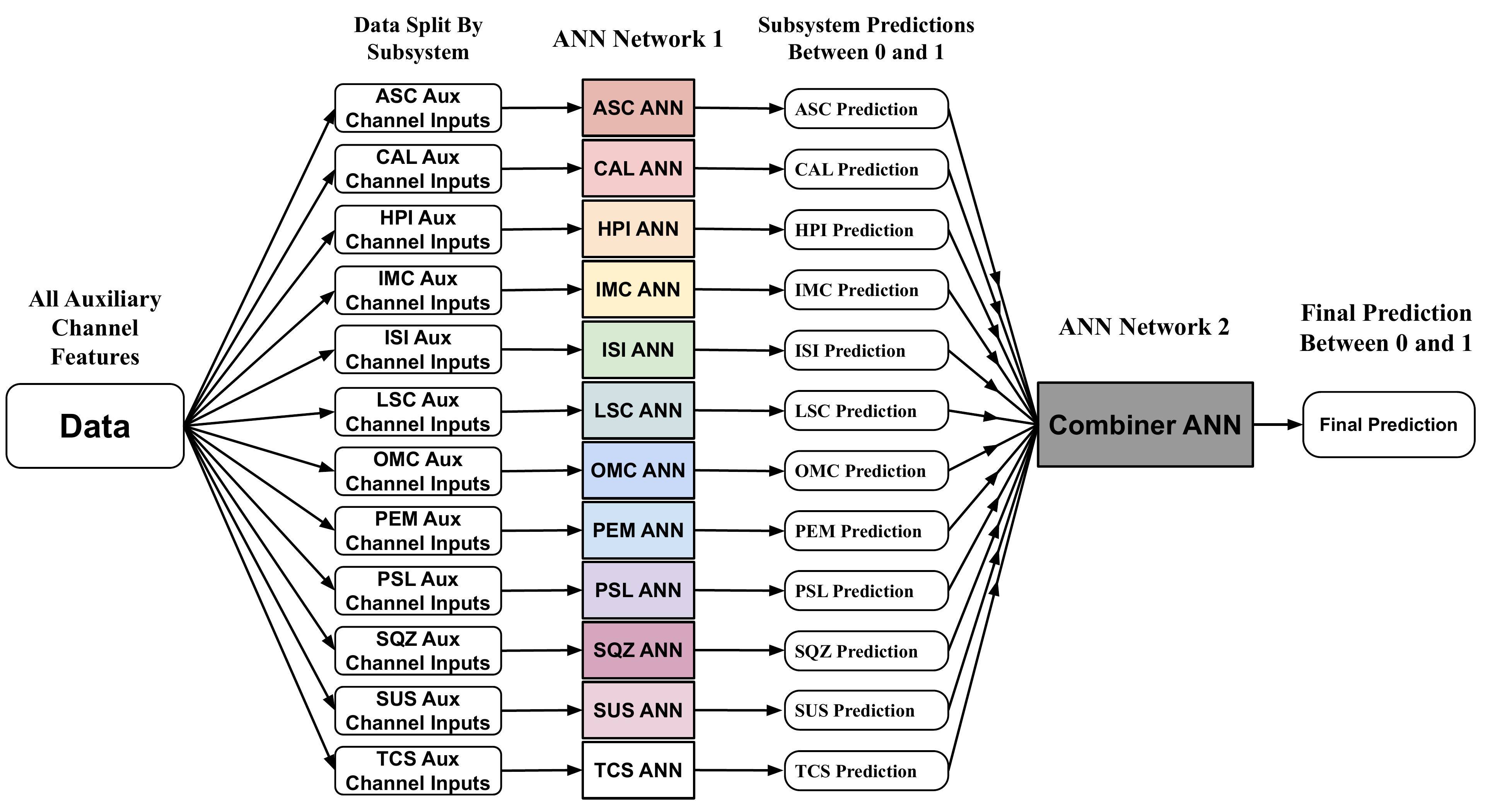}
    \caption{Flowchart of the TITAN workflow.  Auxiliary-channel features are split by subsystem and passed to twelve first-layer ANNs; their predictions are combined by a thirteenth ANN, which outputs a final glitch-classification score between 0 and 1.}
    \label{fig:titanflowchart}
\end{figure*}

TITAN (Tuned Identifier of Transient Auxiliary Noise) is built from twelve neural networks, one for each auxiliary channel subsystem, whose outputs feed a thirteenth combiner network.
This design is an instance of a \emph{mixture-of-experts} architecture, in which each network specializes in one part of the input space and a combiner weighs their predictions.
We built TITAN to test our hypothesis that a mixture-of-experts architecture will improve the performance of ANNs on this problem.
In TITAN's first layer, each of the twelve neural networks receives features from only one auxiliary channel subsystem.
Restricting a network to a single subsystem reduces its number of inputs significantly.
For instance, the ASC subsystem consists of 78 auxiliary channels, so the ASC network has 780 inputs rather than the full 8,440.
Because auxiliary channel subsystems are organized by detector function, each first-layer network can specialize as an expert for its subsystem.
If the subsystems that dominate glitch production change over time, this specialization could help TITAN generalize better over time.
Like GIANTS, each subsystem network gives a prediction between 0 and 1.
The combiner aggregates these twelve scores into a final glitch-classification score for the event.
TITAN can be visualized by the simple workflow presented in Fig.~\ref{fig:titanflowchart}.

TITAN uses the same 90\%/10\% train/test partition introduced in Sec.~\ref{sec:training}: the 2,718 held-out events form the test set, and the remaining 90\% of the dataset is reserved for training.
Because TITAN has two layers, we subdivide this 90\% training pool into two disjoint subsets, one for each layer.
Two-thirds of the pool (60\% of the full dataset, 16,309 events) trains the twelve first-layer networks, and the remaining one-third (30\%, 8,155 events) trains the combiner.
As described in Sec.~\ref{sec:training}, each network further reserves 10\% of its assigned subset as a validation set for early stopping and checkpoint selection.
In absolute terms, the first-layer networks receive 54\% (14,678 events) for training and 6\% (1,631 events) for validation, while the combiner receives 27\% (7,339 events) for training and 3\% (815 events) for validation.
The twelve first-layer networks are trained in parallel on the first subset; once trained, their predictions on the second subset serve as the inputs that train the combiner.
Both layers are also hyperparameter tuned, and App.~\ref{sec:appendix_b} lists the tuned configuration for each of TITAN's ANNs.

\begin{figure}[h!]
    \centering
    \includegraphics[width=1.0\linewidth]{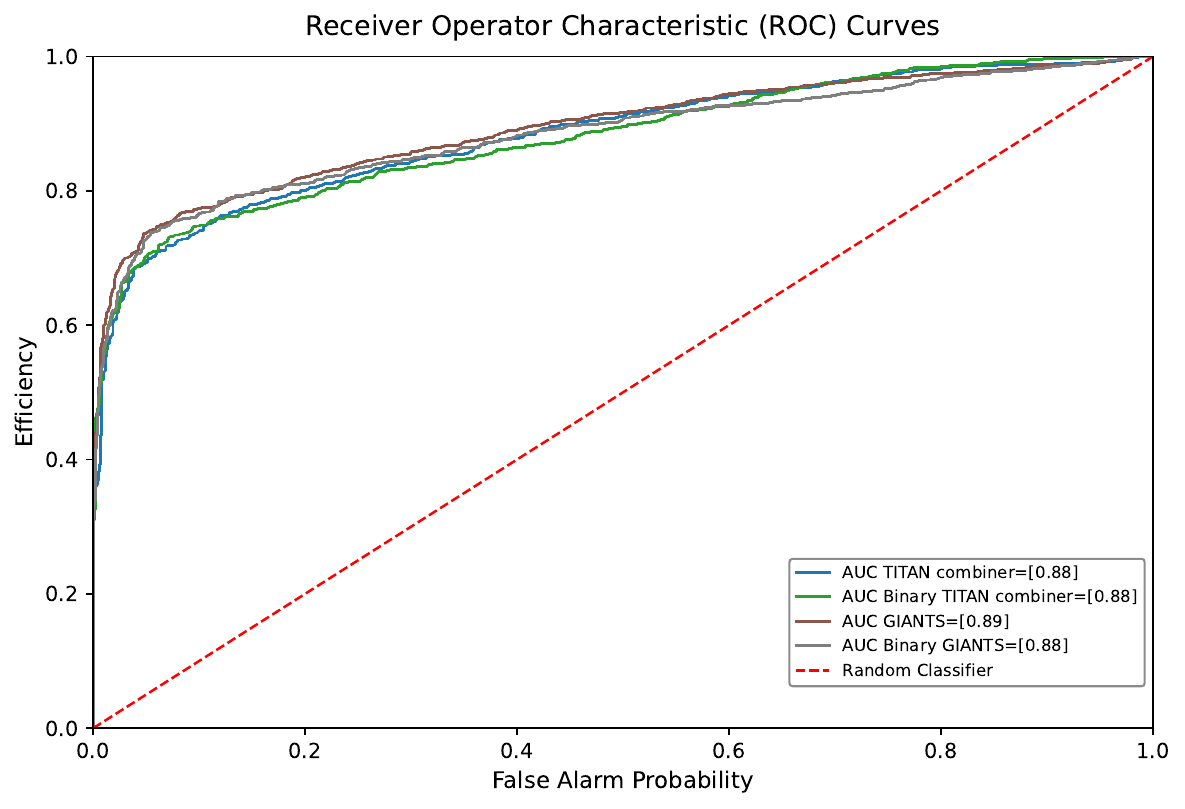}
    \caption{ROC curves for GIANTS and the TITAN combiner on the 10\% held-out test set drawn from the four-week O3 training period, together with their binary-feature variants introduced in Sec.~\ref{sec:features}. Efficiency (true-positive rate) is plotted against false-alarm probability (false-positive rate); the AUC of each curve is given in the legend. All four models achieve nearly identical performance on this test set, and the random-classifier diagonal is shown for reference.}
    \label{fig:ROC}
\end{figure}

To establish a baseline, we trained and tested both GIANTS and TITAN on the four-week O3 dataset described in Sec.~\ref{sec:training}, scoring each model by the AUC of its ROC curve on the 10\% held-out test set. 
Fig.~\ref{fig:ROC} overlays the resulting ROC curves. 
Despite their different architectures, GIANTS and TITAN achieve almost identical AUC scores. 
(Fig.~\ref{fig:ROC} also previews two ``binary" variants of these models, which we introduce in Sec.~\ref{sec:features}.)

\begin{figure}[h!]
    \centering
    \includegraphics[width=1.0\linewidth]{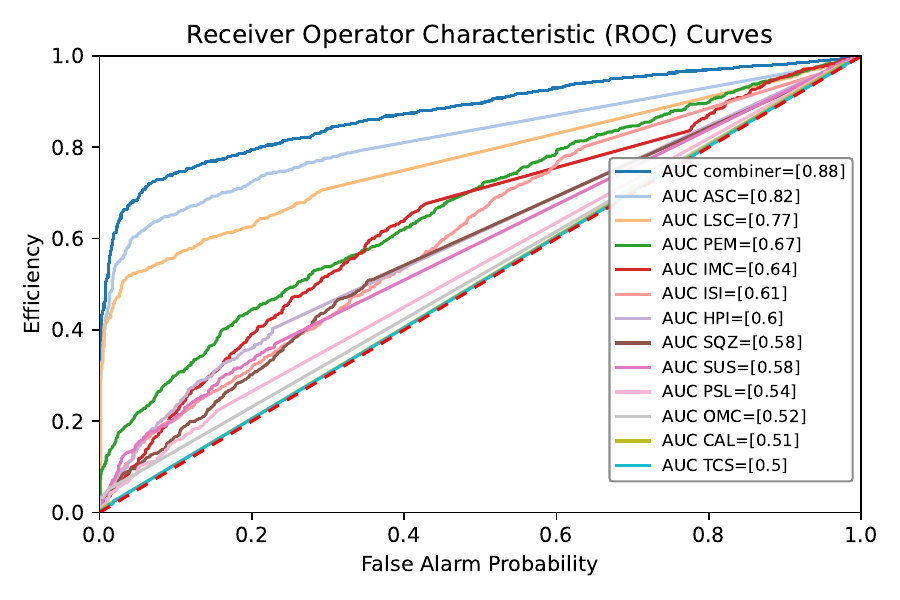}
    \caption{ROC curves for each of TITAN's twelve first-layer subsystem ANNs and for the combiner ANN, evaluated on the same 10\% held-out test set as Fig.~\ref{fig:ROC}. Efficiency is plotted against false-alarm probability; the AUC of each curve is given in the legend. The combiner achieves a higher AUC than any individual subsystem network.}
    \label{fig:TITANROC}
\end{figure}

Fig.~\ref{fig:TITANROC} breaks TITAN's result down by subsystem, plotting the ROC curve of each first-layer ANN alongside the combiner. 
The combiner ($\mathrm{AUC}=0.88$) outperforms every individual subsystem network, consistent with TITAN's design: no single subsystem captures the full glitch population on its own, and the combiner aggregates their partial views into a stronger classifier. Having shown that GIANTS and TITAN perform comparably on data drawn from the training period, we next examine which auxiliary-channel features actually carry predictive information (Sec.~\ref{sec:features}) and how each model holds up on data from other stretches of O3 (Sec.~\ref{sec:ovl_comp}).

\section{Feature Study}
\label{sec:features}
In addition to revisiting the use of ANNs for implementation in iDQ, we also revisited a feature study like the one presented in Ref.~\cite{Biswas_2013}.
Our feature study looked at the significance of each feature separately before considering combinations of features that might together provide boosts in performance.

\begin{figure}[h!]
    \centering
    \includegraphics[width=1.0\linewidth]{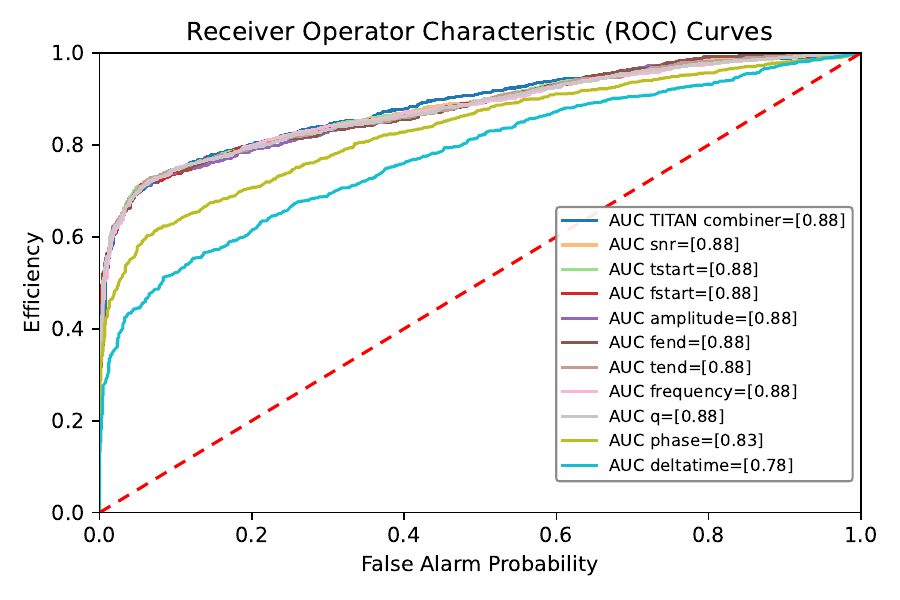}
    \caption{ROC curves for TITAN trained on each trigger feature individually, evaluated on the same 10\% held-out test set as Fig.~\ref{fig:ROC}. Efficiency is plotted against false-alarm probability; the AUC of each single-feature variant is given in the legend. Eight of the ten features yield nearly identical performance, while \texttt{deltatime} and \texttt{phase} produce noticeably weaker classifiers.}
    \label{fig:featstudyroc}
\end{figure}

Fig.~\ref{fig:featstudyroc} presents ROC curves for TITAN trained and tested using the single trigger features labeled in the legend.
Outside of the features \texttt{deltatime} and \texttt{phase}, the other eight features all return nearly identical ROC curves to each other as well as to the combined feature result.
This suggests that, although trigger generators like Omicron provide a wealth of information about auxiliary channel activity, most of this information is not useful in training our models.
This also suggests that we do not need to consider subsets of feature combinations in this study. 

The uniform performance across individual features suggested that feature values carry little predictive information and that the MLAs learn primarily from whether auxiliary-channel activity exceeded the SNR threshold.
To test this hypothesis, we constructed a binary feature set in which each of the 844 auxiliary channels is represented by a single value: one if the channel exhibited activity above threshold within the window of interest and zero otherwise.
We then trained and hyperparameter-tuned TITAN and GIANTS on this set, yielding models we refer to as ``binary-TITAN'' and ``binary-GIANTS".  
We saw that our binary feature set produced a nearly identical ROC curve when evaluated on the held-out test dataset, as seen in Fig.~\ref{fig:ROC}.

\section{Model Generalization}
\label{sec:ovl_comp}
While strong performance on a held-out test set drawn from the training period establishes that GIANTS and TITAN are viable algorithms for this classification problem, this result does not indicate how either model behaves on data collected outside the training period.
To assess generalization of these algorithms, we evaluated GIANTS, TITAN, their binary counterparts, and the OVL classifier on the ten independent test sets introduced in Tab.~\ref{tab: test_list}.
The variation of size and time in each test set allowed us to view the strength of our training set and the corresponding model's generalization across time.
Each model was trained on the same four-week training set used throughout this work, allowing for a direct comparison between OVL and our MLAs. 
The dependence of these results on the choice of training data is examined in App.~\ref{sec:appendix_a}.

\begin{figure*}[htbp]
    \centering
    \includegraphics[width=\textwidth]{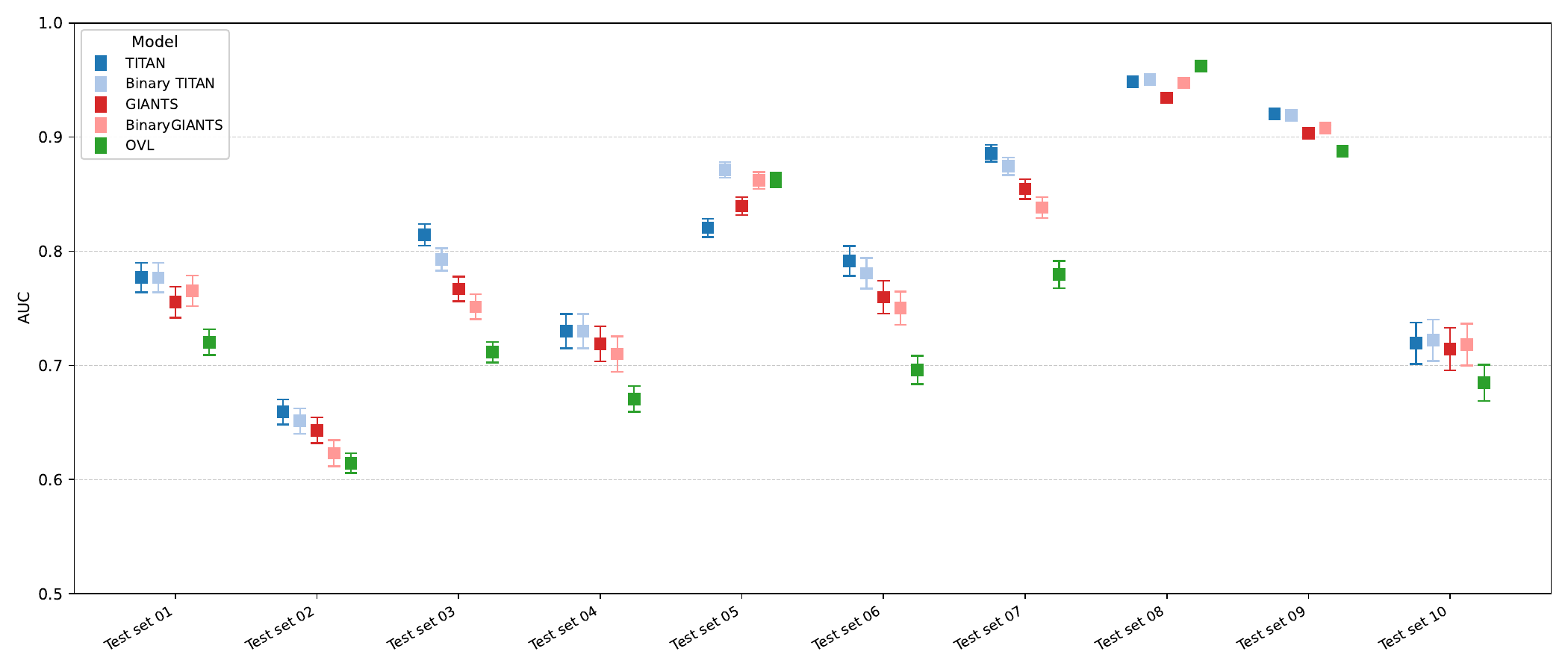}
    \caption{AUC achieved by TITAN, binary-TITAN, GIANTS, binary-GIANTS, and OVL on the ten test sets of Tab.~\ref{tab: test_list}. Error bars are derived from the variance estimator of DeLong et al.~\cite{DeLong1988}. Binary-TITAN outperforms OVL on nine of the ten test sets, TITAN and binary-GIANTS on eight, and GIANTS on seven.  TITAN's AUC varies less from set to set than GIANTS's, while the binary variants track their full-feature counterparts across the run.}
    \label{fig:TestROCcomp}
\end{figure*}

Fig.~\ref{fig:TestROCcomp} reports the AUC achieved by each model on each test set. 
Binary-TITAN outperforms OVL on nine of the ten test sets, TITAN and binary-GIANTS each on eight, and GIANTS on seven. 
Here a model outperforms OVL when its AUC exceeds OVL’s by more than the combined uncertainty. 
OVL achieves the highest AUC of any model only on test set 08.
TITAN further delivers more consistent performance than GIANTS across the observing run, with smaller set-to-set variation in AUC.
This is consistent with the mixture-of-experts design introduced in Sec.~\ref{sec:models}.  
When the dominant sources of glitches shift between the training period and a test set, the degradation is confined to the affected subsystem networks rather than spread across a single monolithic model, allowing the combiner to lean on subsystems whose predictive power is preserved.

The binary variants of both models track their full-feature counterparts.
Individual test sets show differences of either sign, the largest being test set 05, where binary-TITAN exceeds TITAN by about 0.05 in AUC, but averaged over the ten test sets the two representations are nearly indistinguishable: 0.808 for both TITAN and binary-TITAN, and 0.793 versus 0.789 for GIANTS and binary-GIANTS (Tab.~\ref{tab:trainstems}).
The binary representation therefore incurs no systematic loss in performance, indicating that the predictive content of an Omicron feature vector is carried almost entirely by the binary indicator (cf.\ Sec.~\ref{sec:features}).
Because the binary representation reduces the per-sample input dimensionality by roughly an order of magnitude, binary-TITAN and binary-GIANTS are the practical choices for deployment in iDQ.

Together, these results demonstrate that an ANN trained on auxiliary-channel features outperforms OVL across diverse stretches of O3, with TITAN providing the most consistent generalization across the run.

\section{Discussion}
\label{sec:discussion}
Throughout this work, we tested multiple ANN architectures.
These included GIANTS, a single neural network that sees all auxiliary channels at once, and TITAN, a mixture-of-experts model built from twelve subsystem-specific networks combined by a thirteenth.
Both architectures were hyperparameter tuned.
On the held-out test set drawn from the four-week training period, there was no noticeable difference in AUC between the architectures.

However, on test data drawn from other stretches of O3, the difference between TITAN and GIANTS was noticeable.
TITAN generalized more consistently across time than GIANTS, and every model outperformed OVL on at least seven of the ten test sets (Fig.~\ref{fig:TestROCcomp}), with binary-TITAN doing so on nine.
This supports our hypothesis that the subsystem-specific networks in TITAN allow for better generalization.

Our feature study addressed a separate question: which Omicron trigger features carry predictive information. 
When we removed all but one feature, TITAN achieved nearly the same AUC for eight of the ten features, which suggested that the networks learn primarily from whether a channel was active and not from the feature values themselves. 
Training on a binary indicator of channel activity confirmed this. 
With the binary format, there was no systematic decrease in AUC across the ten test sets for either architecture. 
We conclude that the presence of a trigger in an auxiliary channel carries nearly all of the predictive information, and that the binary representation reduces the input dimensionality tenfold without sacrificing performance.
The binary models are also markedly less sensitive to the choice of training data (App.~\ref{sec:appendix_a}).

Historically, OVL's performance has been comparable to that of earlier single-network ANN approaches.
With recent updates to TensorFlow and the ability to perform hyperparameter tuning, ANNs of this kind now show improved performance.
When trained on the same data, not only does GIANTS outperform OVL, but TITAN's mixture-of-experts design generalizes even better to unseen test periods.

Two directions for future work extend the present study.
First, we plan to integrate GIANTS and TITAN, particularly their binary variants, whose tenfold smaller input makes them well suited to low-latency operation, into the modern iDQ infrastructure.
Second, we plan to explore a new mixture-of-experts architecture in which each expert is trained on a specific class of glitch rather than a specific detector subsystem, using Gravity Spy \cite{Zevin:2016qwy, Zevin:2023rmt, Glanzer:2022avx} classifications as labels.
Such a model would address two gaps in the present work.
First, because Gravity Spy provides categorical labels, the resulting MLA could predict the type of glitch present based on auxiliary activity only rather than  a binary glitch/clean label.
Second, because its experts would learn glitch morphologies rather than subsystem signatures, the model may prove more robust to the non-stationarities that limit our current models.
Together, these directions chart a path from the long-standing OVL-based classifier toward an ANN-based iDQ pipeline that is more consistent across time and able to provide real-time diagnostic information about individual glitches.

\acknowledgements
The authors are grateful for Christopher Berry's helpful review and comments on the initial draft of this work. 

The authors are grateful to the LVK for access to the data used to train and test the models in this work.  The training and testing data was derived from Omicron triggers produced by the LVK \cite{Robinet_2020}.

This material is based upon work supported by NSF's LIGO Laboratory which is a major facility fully funded by the National Science Foundation (NSF).
LIGO was constructed by the California Institute of Technology and Massachusetts Institute of Technology with funding from the United States NSF, and operates under cooperative agreement PHY-2309200. Advanced LIGO was built under award PHY-0823459.
The authors gratefully acknowledge the support of the United States NSF for the construction and operation of the LIGO Laboratory and Advanced LIGO as well as the Science
and Technology Facilities Council (STFC) of the United Kingdom, the Max-Planck-Society
(MPS), and the State of Niedersachsen/Germany for support of the construction of Advanced LIGO and construction and operation of the GEO600 detector. 
Additional support for Advanced LIGO was provided by the Australian Research Council (ARC).
The authors are grateful for computational resources provided by the LIGO Laboratory and supported by NSF Grants PHY-0757058 and PHY-0823459.
The authors are supported by NSF grant PHY-2308796.  

\section*{Data Availability}
The code used to perform the analyses in this work is publicly 
available at \url{https://gitlab.com/madeline.wade/ligo-noise-mla/-/tags/v1.0.0}, Ref.~\cite{ligo_noise_mla_code}. The datasets used in this study are available at \url{https://zenodo.org/records/23692820}, Ref.~\cite{data_release}.

\appendix
\section{Dependence of Model Performance on Training Data}
\label{sec:appendix_a}
\begin{figure*}
  \centering
  \includegraphics[width=\textwidth]{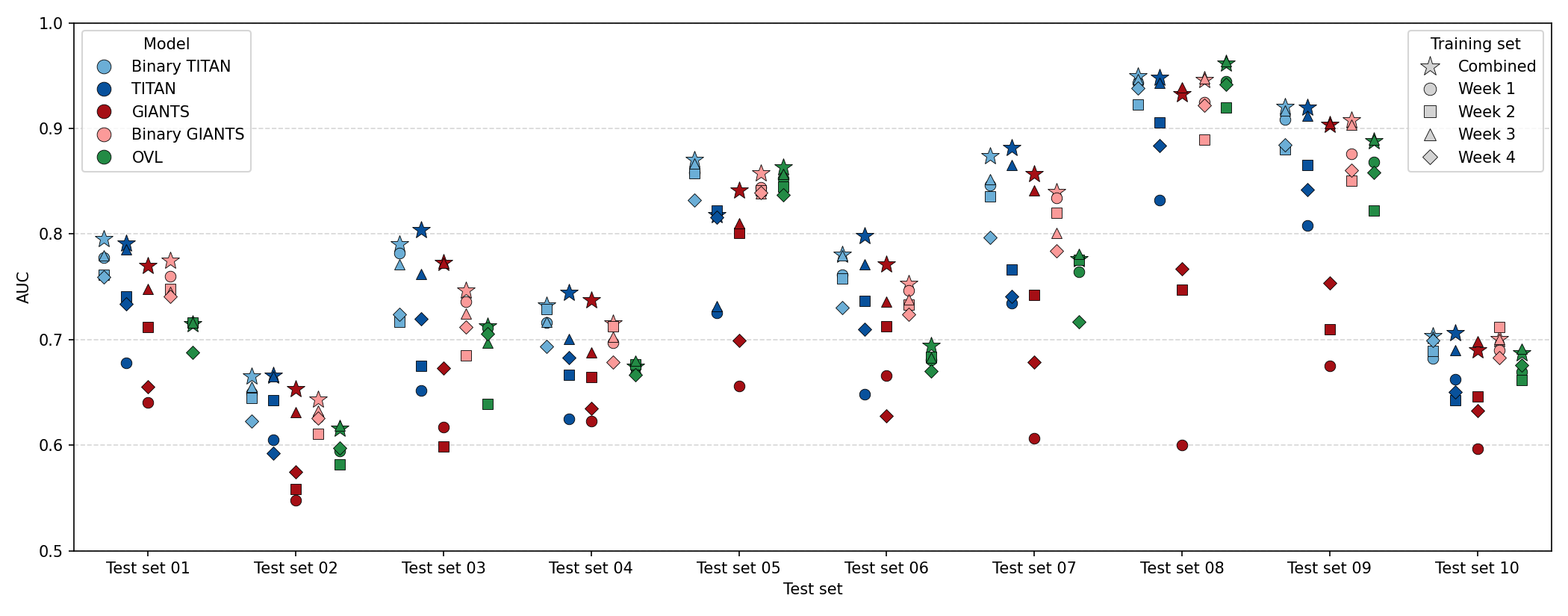}
  \caption{AUC achieved by binary-TITAN, TITAN, GIANTS, binary-GIANTS, and OVL on the ten independent test sets of Tab.~\ref{tab: test_list}, for five training configurations: the combined four-week set used throughout the main text and each of the four single-week intervals of Tab.~\ref{tab:clustering-counts-train} in
  isolation. Marker color denotes the model and marker shape the training set (star: combined; circle, square, triangle, diamond: Weeks 1--4); each cluster along the horizontal axis is one test set. The full-feature models lose substantial AUC when trained on a single week, whereas the binary models and OVL remain comparatively stable.}
  \label{fig:trainstems}
  \end{figure*}

To assess how strongly each classifier depends on the breadth of its training data, we retrained every model on each of the four disjoint single-week intervals of Tab.~\ref{tab:clustering-counts-train} in isolation and compared the result to the combined four-week training set used throughout the main text.
Each of the five resulting configurations per model---four single-week and one combined---was evaluated on the same ten independent test sets of Tab.~\ref{tab: test_list}.
Fig.~\ref{fig:trainstems} shows the per-test-set AUC for all five models and five training configurations, and Tab.~\ref{tab:trainstems} reports the mean AUC over the ten test sets.

For all four ANNs, combined training meets or exceeds every single-week configuration: binary-TITAN, TITAN, GIANTS, and binary-GIANTS each attain their highest mean AUC on the full four-week set.
The effect is largest for the full-feature models, whose single-week mean AUC falls as much as $0.17$ below their combined value.
By contrast, OVL gains little from the combined set.  Its combined AUC ($0.759$) barely edges its best single week ($0.758$), which reflects that OVL's channel ranking saturates quickly and does not exploit additional training examples the way a trained network does.

The binary representations are markedly more robust to the choice of training set.
Across the four single-week intervals, binary-TITAN's mean AUC spans only $0.031$ and binary-GIANTS' only $0.017$, whereas full-feature GIANTS swings $0.154$ and full-feature TITAN $0.086$.
Reducing each channel to a single presence indicator thus not only lowers the input dimensionality by roughly an order of magnitude but also makes the classifiers far less sensitive to how much training data is available, which is an attractive property for a detector-characterization pipeline that is periodically retrained as instrument behavior evolves.

Finally, when a model is trained on only a single week, the classical OVL baseline becomes competitive with, and on the least favorable weeks superior to, the full-feature ANNs.
Full-feature GIANTS falls to $0.623$ when trained on Week~1, below OVL's lowest single-week AUC of $0.732$;
every one of OVL's single-week configurations ($0.732$--$0.758$) outperforms GIANTS trained on Week~1.
OVL's stability across training sets (range $0.026$) mirrors that of the binary ANNs, and underscores that the advantage of the full-feature networks is realized only once sufficient training data is available.

Taken together, these trends reinforce the architectural choices of the main text.
The combined training set is preferred wherever data volume permits, and the binary representations provide a robust fallback when it does not.

  \begin{table*}
  \centering
  \begin{tabular}{l c c c c c c}
  \hline\hline
  Model & Combined & Week 1 & Week 2 & Week 3 & Week 4 & Range \\
  \hline
  Binary TITAN  & 0.808 & 0.793 & 0.780 & 0.799 & 0.768 & 0.031 \\
  TITAN         & 0.808 & 0.697 & 0.746 & 0.783 & 0.737 & 0.086 \\
  GIANTS        & 0.793 & 0.623 & 0.689 & 0.777 & 0.670 & 0.154 \\
  Binary GIANTS & 0.789 & 0.774 & 0.760 & 0.774 & 0.757 & 0.017 \\
  OVL           & 0.759 & 0.748 & 0.732 & 0.758 & 0.736 & 0.026 \\
  \hline\hline
  \end{tabular}
  \caption{Mean AUC over the ten independent test sets of Tab.~\ref{tab: test_list} for each model and training configuration. The final column gives the spread (maximum minus minimum) across the four single-week intervals. Combined training meets or exceeds every single-week configuration for all four ANNs, and the binary representations vary far less across training sets than their full-feature counterparts.}
  \label{tab:trainstems}
  \end{table*}

\section{Hyperparameters}
\label{sec:appendix_b}
This appendix lists the tuned hyperparameter configurations for each model presented in this work.
All values were selected using the Keras Bayesian Optimization Tuner over the search ranges given in Tab.~\ref{tab:params}, following the protocol described in Sec.~\ref{sec:hyper-tuning}: 100 trials per model, up to 1,000 epochs per trial, with validation loss as the optimization objective.
For TITAN, each of the twelve first-layer subsystem networks and the combiner network were tuned independently.
The one exception is binary-TITAN, whose hyperparameters were selected by manual tuning; the Bayesian search did not yield a configuration that outperformed the manual baseline on either the four-week or ten independent test sets, so we report the manual configuration here.

\begin{table*}[p]
\centering
\small
\renewcommand{\arraystretch}{1.2}
\begin{tabular}{|l|l|l|}
\hline
\textbf{Parameter} & \textbf{GIANTS} & \textbf{Binary GIANTS} \\
\hline
Layers & 3 & 2 \\
\hline
Nodes (per layer) & [224, 32, 32] & [256, 64] \\
\hline
Activation (per layer) & [`softplus', `tanh', `softsign'] & [`linear', `gelu'] \\
\hline
Dropout (per layer) & [0.0, 0.2, 0.3] & [0.3, 0.1] \\
\hline
L1 weight (per layer) & [0.001, 0.01, 0.0] & [0.0, 0.001] \\
\hline
L2 weight (per layer) & [0.01, 0.0, 0.0001] & [0.001, 0.001] \\
\hline
Epochs & 6000 & 6000 \\
\hline
Batch size & 16 & 16 \\
\hline
Optimizer & SGD & RMSprop \\
\hline
Learning rate & 0.01 & 0.001 \\
\hline
Loss & binary\_crossentropy & binary\_crossentropy \\
\hline
Metric & accuracy & accuracy \\
\hline
Early stopping monitor & val\_loss & val\_loss \\
\hline
Early stopping min\_delta & 1.0$\times 10^{-4}$ & 1.0$\times 10^{-4}$ \\
\hline
Early stopping patience & 15 & 15 \\
\hline
Early stopping mode & auto & auto \\
\hline
\end{tabular}
\caption{Comparison of hyper-tuned configurations used to train the GIANTS and Binary GIANTS MLAs (values shown per layer where applicable).}
\label{tab:giants_vs_binary_giants}
\end{table*}

\begin{table*}[p]
\centering
\small
\renewcommand{\arraystretch}{1.2}
\begin{tabular}{|l|l|l|}
\hline
\textbf{Parameter} & \textbf{ISI} & \textbf{Binary ISI} \\
\hline
Layers & 2 & 1 \\
\hline
Nodes (per layer) & [32, 1] & [109] \\
\hline
Activation (per layer) & [`softsign', `sigmoid'] & [`tanh'] \\
\hline
Dropout (per layer) & [0.4, 0.0] & [0.0] \\
\hline
L1 weight (per layer) & [0.001, 0.0] & [0.0001] \\
\hline
L2 weight (per layer) & [0.0, 0.0] & [0.0001] \\
\hline
Epochs & 6000 & 500 \\
\hline
Batch size & 16 & 16 \\
\hline
Optimizer & Nadam & RMSprop \\
\hline
Learning rate & 0.0001 & 0.001 \\
\hline
Loss & binary\_crossentropy & binary\_crossentropy \\
\hline
Metric & accuracy & accuracy \\
\hline
Early stopping monitor & val\_loss & val\_loss \\
\hline
Early stopping min\_delta & 1.0$\times 10^{-4}$ & 1.0$\times 10^{-4}$ \\
\hline
Early stopping patience & 15 & 15 \\
\hline
Early stopping mode & auto & auto \\
\hline
\end{tabular}
\caption{Comparison of hyper-tuned and binary-training configurations for the ISI subsystem ANN (values shown per layer where applicable).}
\label{tab:isi_vs_binary_isi}
\end{table*}

\begin{table*}[p]
\centering
\small
\renewcommand{\arraystretch}{1.2}
\begin{tabular}{|l|l|l|}
\hline
\textbf{Parameter} & \textbf{LSC} & \textbf{Binary LSC} \\
\hline
Layers & 4 & 1 \\
\hline
Nodes (per layer) & [192, 320, 320, 96] & [42] \\
\hline
Activation (per layer) & [`softsign', `softsign', `relu', `relu'] & [`tanh'] \\
\hline
Dropout (per layer) & [0.2, 0.7, 0.2, 0.7] & [0.0] \\
\hline
L1 weight (per layer) & [0.0001, 0.0, 0.01, 0.0] & [0.0001] \\
\hline
L2 weight (per layer) & [0.0001, 0.0001, 0.0, 0.0001] & [0.0001] \\
\hline
Epochs & 6000 & 500 \\
\hline
Batch size & 16 & 16 \\
\hline
Optimizer & SGD & RMSprop \\
\hline
Learning rate & 0.001 & 0.001 \\
\hline
Loss & binary\_crossentropy & binary\_crossentropy \\
\hline
Metric & accuracy & accuracy \\
\hline
Early stopping monitor & val\_loss & val\_loss \\
\hline
Early stopping min\_delta & 1.0$\times 10^{-4}$ & 1.0$\times 10^{-4}$ \\
\hline
Early stopping patience & 15 & 15 \\
\hline
Early stopping mode & auto & auto \\
\hline
\end{tabular}
\caption{Comparison of hyper-tuned and binary-training configurations for the LSC subsystem ANN (values shown per layer where applicable).}
\label{tab:lsc_vs_binary_lsc}
\end{table*}

\begin{table*}[p]
\centering
\small
\renewcommand{\arraystretch}{1.2}
\begin{tabular}{|l|l|l|}
\hline
\textbf{Parameter} & \textbf{HPI} & \textbf{Binary HPI} \\
\hline
Layers & 2 & 2 \\
\hline
Nodes (per layer) & [480, 256] & [82, 342] \\
\hline
Activation (per layer) & [`softplus', `relu'] & [`softsign', `relu'] \\
\hline
Dropout (per layer) & [0.6, 0.4] & [0.6, 0.0] \\
\hline
L1 weight (per layer) & [0.0, 0.01] & [0.0, 0.0] \\
\hline
L2 weight (per layer) & [0.01, 0.01] & [0.0001, 0.01] \\
\hline
Epochs & 6000 & 500 \\
\hline
Batch size & 16 & 16 \\
\hline
Optimizer & SGD & RMSprop \\
\hline
Learning rate & 0.001 & 0.001 \\
\hline
Loss & binary\_crossentropy & binary\_crossentropy \\
\hline
Metric & accuracy & accuracy \\
\hline
Early stopping monitor & val\_loss & val\_loss \\
\hline
Early stopping min\_delta & 1.0$\times 10^{-4}$ & 1.0$\times 10^{-4}$ \\
\hline
Early stopping patience & 15 & 15 \\
\hline
Early stopping mode & auto & auto \\
\hline
\end{tabular}
\caption{Comparison of hyper-tuned and binary-training configurations for the HPI subsystem ANN (values shown per layer where applicable).}
\label{tab:hpi_vs_binary_hpi}
\end{table*}

\begin{table*}[p]
\centering
\small
\renewcommand{\arraystretch}{1.2}
\begin{tabular}{|l|l|l|}
\hline
\textbf{Parameter} & \textbf{ASC} & \textbf{Binary ASC} \\
\hline
Layers & 3 & 2 \\
\hline
Nodes (per layer) & [64, 384, 160] & [2, 602] \\
\hline
Activation (per layer) & [`relu', `linear', `selu'] & [`swish', `tanh'] \\
\hline
Dropout (per layer) & [0.1, 0.2, 0.5] & [0.7, 0.2] \\
\hline
L1 weight (per layer) & [0.0, 0.0, 0.001] & [0.0, 0.0] \\
\hline
L2 weight (per layer) & [0.01, 0.001, 0.0] & [0.0001, 0.001] \\
\hline
Epochs & 6000 & 500 \\
\hline
Batch size & 16 & 16 \\
\hline
Optimizer & RMSprop & Adam \\
\hline
Learning rate & 0.01 & 0.0001 \\
\hline
Loss & binary\_crossentropy & binary\_crossentropy \\
\hline
Metric & accuracy & accuracy \\
\hline
Early stopping monitor & val\_loss & val\_loss \\
\hline
Early stopping min\_delta & 1.0$\times 10^{-4}$ & 1.0$\times 10^{-4}$ \\
\hline
Early stopping patience & 15 & 15 \\
\hline
Early stopping mode & auto & auto \\
\hline
\end{tabular}
\caption{Comparison of hyper-tuned and binary-training configurations for the ASC subsystem ANN (values shown per layer where applicable).}
\label{tab:asc_vs_binary_asc}
\end{table*}

\begin{table*}[p]
\centering
\small
\renewcommand{\arraystretch}{1.2}
\begin{tabular}{|l|l|l|}
\hline
\textbf{Parameter} & \textbf{SUS} & \textbf{Binary SUS} \\
\hline
Layers & 2 & 1 \\
\hline
Nodes (per layer) & [192, 192] & [682] \\
\hline
Activation (per layer) & [`elu', `swish'] & [`tanh'] \\
\hline
Dropout (per layer) & [0.5, 0.3] & [0.5] \\
\hline
L1 weight (per layer) & [0.0001, 0.0] & [0.0] \\
\hline
L2 weight (per layer) & [0.0, 0.01] & [0.0001] \\
\hline
Epochs & 6000 & 500 \\
\hline
Batch size & 16 & 16 \\
\hline
Optimizer & RMSprop & RMSprop \\
\hline
Learning rate & 0.01 & 0.0001 \\
\hline
Loss & binary\_crossentropy & binary\_crossentropy \\
\hline
Metric & accuracy & accuracy \\
\hline
Early stopping monitor & val\_loss & val\_loss \\
\hline
Early stopping min\_delta & 1.0$\times 10^{-4}$ & 1.0$\times 10^{-4}$ \\
\hline
Early stopping patience & 15 & 15 \\
\hline
Early stopping mode & auto & auto \\
\hline
\end{tabular}
\caption{Comparison of hyper-tuned and binary-training configurations for the SUS subsystem ANN (values shown per layer where applicable).}
\label{tab:sus_vs_binary_sus}
\end{table*}

\begin{table*}[p]
\centering
\small
\renewcommand{\arraystretch}{1.2}
\begin{tabular}{|l|l|l|}
\hline
\textbf{Parameter} & \textbf{SQZ} & \textbf{Binary SQZ} \\
\hline
Layers & 5 & 1 \\
\hline
Nodes (per layer) & [416, 512, 64, 416, 1] & [55] \\
\hline
Activation (per layer) & [`relu', `tanh', `linear', `selu', `sigmoid'] & [`relu'] \\
\hline
Dropout (per layer) & [0.1, 0.2, 0.8, 0.0, 0.0] & [0.0] \\
\hline
L1 weight (per layer) & [0.0, 0.0, 0.0, 0.0, 0.0] & [0.0] \\
\hline
L2 weight (per layer) & [0.0001, 0.001, 0.001, 0.0001, 0.0] & [0.001] \\
\hline
Epochs & 6000 & 500 \\
\hline
Batch size & 16 & 16 \\
\hline
Optimizer & Nadam & Adamax \\
\hline
Learning rate & 0.01 & 0.01 \\
\hline
Loss & binary\_crossentropy & binary\_crossentropy \\
\hline
Metric & accuracy & accuracy \\
\hline
Early stopping monitor & val\_loss & val\_loss \\
\hline
Early stopping min\_delta & 1.0$\times 10^{-4}$ & 1.0$\times 10^{-4}$ \\
\hline
Early stopping patience & 15 & 15 \\
\hline
Early stopping mode & auto & auto \\
\hline
\end{tabular}
\caption{Comparison of hyper-tuned and binary-training configurations for the SQZ subsystem ANN (values shown per layer where applicable).}
\label{tab:sqz_vs_binary_sqz}
\end{table*}

\begin{table*}[p]
\centering
\small
\renewcommand{\arraystretch}{1.2}
\begin{tabular}{|l|l|l|}
\hline
\textbf{Parameter} & \textbf{PEM} & \textbf{Binary PEM} \\
\hline
Layers & 3 & 1 \\
\hline
Nodes (per layer) & [448, 288, 480] & [2329] \\
\hline
Activation (per layer) & [`selu', `selu', `softplus'] & [`relu'] \\
\hline
Dropout (per layer) & [0.8, 0.7, 0.5] & [0.3] \\
\hline
L1 weight (per layer) & [0.0001, 0.0, 0.0001] & [0.0] \\
\hline
L2 weight (per layer) & [0.0, 0.0001, 0.0] & [0.001] \\
\hline
Epochs & 6000 & 500 \\
\hline
Batch size & 16 & 16 \\
\hline
Optimizer & Nadam & RMSprop \\
\hline
Learning rate & 0.01 & 0.001 \\
\hline
Loss & binary\_crossentropy & binary\_crossentropy \\
\hline
Metric & accuracy & accuracy \\
\hline
Early stopping monitor & val\_loss & val\_loss \\
\hline
Early stopping min\_delta & 1.0$\times 10^{-4}$ & 1.0$\times 10^{-4}$ \\
\hline
Early stopping patience & 15 & 15 \\
\hline
Early stopping mode & auto & auto \\
\hline
\end{tabular}
\caption{Comparison of hyper-tuned and binary-training configurations for the PEM subsystem ANN (values shown per layer where applicable).}
\label{tab:pem_vs_binary_pem}
\end{table*}

\begin{table*}[p]
\centering
\small
\renewcommand{\arraystretch}{1.2}
\begin{tabular}{|l|l|l|}
\hline
\textbf{Parameter} & \textbf{IMC} & \textbf{Binary IMC} \\
\hline
Layers & 4 & 2 \\
\hline
Nodes (per layer) & [352, 352, 256, 64] & [247, 37] \\
\hline
Activation (per layer) & [`softplus', `elu', `tanh', `selu'] & [`softsign', `relu'] \\
\hline
Dropout (per layer) & [0.2, 0.2, 0.0, 0.0] & [0.2, 0.1] \\
\hline
L1 weight (per layer) & [0.0001, 0.0, 0.0, 0.0001] & [0.0, 0.0001] \\
\hline
L2 weight (per layer) & [0.0001, 0.0001, 0.001, 0.0] & [0.0001, 0.0] \\
\hline
Epochs & 6000 & 500 \\
\hline
Batch size & 16 & 16 \\
\hline
Optimizer & Adamax & RMSprop \\
\hline
Learning rate & 0.01 & 0.001 \\
\hline
Loss & binary\_crossentropy & binary\_crossentropy \\
\hline
Metric & accuracy & accuracy \\
\hline
Early stopping monitor & val\_loss & val\_loss \\
\hline
Early stopping min\_delta & 1.0$\times 10^{-4}$ & 1.0$\times 10^{-4}$ \\
\hline
Early stopping patience & 15 & 15 \\
\hline
Early stopping mode & auto & auto \\
\hline
\end{tabular}
\caption{Comparison of hyper-tuned and binary-training configurations for the IMC subsystem ANN (values shown per layer where applicable).}
\label{tab:imc_vs_binary_imc}
\end{table*}

\begin{table*}[p]
\centering
\small
\renewcommand{\arraystretch}{1.2}
\begin{tabular}{|l|l|l|}
\hline
\textbf{Parameter} & \textbf{PSL} & \textbf{Binary PSL} \\
\hline
Layers & 5 & 1 \\
\hline
Nodes (per layer) & [352, 224, 320, 288, 1] & [22] \\
\hline
Activation (per layer) & [`swish', `softplus', `selu', `linear', `sigmoid'] & [`relu'] \\
\hline
Dropout (per layer) & [0.5, 0.3, 0.2, 0.3, 0.0] & [0.0] \\
\hline
L1 weight (per layer) & [0.001, 0.0001, 0.0001, 0.0001, 0.0] & [0.0] \\
\hline
L2 weight (per layer) & [0.0, 0.0, 0.0, 0.0, 0.0] & [0.001] \\
\hline
Epochs & 6000 & 500 \\
\hline
Batch size & 16 & 16 \\
\hline
Optimizer & Nadam & Adamax \\
\hline
Learning rate & 0.01 & 0.01 \\
\hline
Loss & binary\_crossentropy & binary\_crossentropy \\
\hline
Metric & accuracy & accuracy \\
\hline
Early stopping monitor & val\_loss & val\_loss \\
\hline
Early stopping min\_delta & 1.0$\times 10^{-4}$ & 1.0$\times 10^{-4}$ \\
\hline
Early stopping patience & 15 & 15 \\
\hline
Early stopping mode & auto & auto \\
\hline
\end{tabular}
\caption{Comparison of hyper-tuned and binary-training configurations for the PSL subsystem ANN (values shown per layer where applicable).}
\label{tab:psl_vs_binary_psl}
\end{table*}

\begin{table*}[p]
\centering
\small
\renewcommand{\arraystretch}{1.2}
\begin{tabular}{|l|l|l|}
\hline
\textbf{Parameter} & \textbf{OMC} & \textbf{Binary OMC} \\
\hline
Layers & 2 & 1 \\
\hline
Nodes (per layer) & [224, 224] & [2] \\
\hline
Activation (per layer) & [`softplus', `softplus'] & [`softsign'] \\
\hline
Dropout (per layer) & [0.3, 0.4] & [0.0] \\
\hline
L1 weight (per layer) & [0.0, 0.0] & [0.0001] \\
\hline
L2 weight (per layer) & [0.0001, 0.0001] & [0.0] \\
\hline
Epochs & 6000 & 500 \\
\hline
Batch size & 16 & 16 \\
\hline
Optimizer & Adadelta & Nadam \\
\hline
Learning rate & 0.001 & 0.01 \\
\hline
Loss & binary\_crossentropy & binary\_crossentropy \\
\hline
Metric & accuracy & accuracy \\
\hline
Early stopping monitor & val\_loss & val\_loss \\
\hline
Early stopping min\_delta & 1.0$\times 10^{-4}$ & 1.0$\times 10^{-4}$ \\
\hline
Early stopping patience & 15 & 15 \\
\hline
Early stopping mode & auto & auto \\
\hline
\end{tabular}
\caption{Comparison of hyper-tuned and binary-training configurations for the OMC subsystem ANN (values shown per layer where applicable).}
\label{tab:omc_vs_binary_omc}
\end{table*}

\begin{table*}[p]
\centering
\small
\renewcommand{\arraystretch}{1.2}
\begin{tabular}{|l|l|l|}
\hline
\textbf{Parameter} & \textbf{TCS} & \textbf{Binary TCS} \\
\hline
Layers & 4 & 1 \\
\hline
Nodes (per layer) & [448, 448, 320, 320] & [2] \\
\hline
Activation (per layer) & [`softplus', `softplus', `gelu', `elu'] & [`softsign'] \\
\hline
Dropout (per layer) & [0.7, 0.4, 0.1, 0.5] & [0.0] \\
\hline
L1 weight (per layer) & [0.0, 0.0, 0.0001, 0.001] & [0.0001] \\
\hline
L2 weight (per layer) & [0.0001, 0.0001, 0.0001, 0.001] & [0.0] \\
\hline
Epochs & 6000 & 500 \\
\hline
Batch size & 16 & 16 \\
\hline
Optimizer & RMSprop & Nadam \\
\hline
Learning rate & 0.01 & 0.01 \\
\hline
Loss & binary\_crossentropy & binary\_crossentropy \\
\hline
Metric & accuracy & accuracy \\
\hline
Early stopping monitor & val\_loss & val\_loss \\
\hline
Early stopping min\_delta & 1.0$\times 10^{-4}$ & 1.0$\times 10^{-4}$ \\
\hline
Early stopping patience & 15 & 15 \\
\hline
Early stopping mode & auto & auto \\
\hline
\end{tabular}
\caption{Comparison of hyper-tuned and binary-training configurations for the TCS subsystem ANN (values shown per layer where applicable).}
\label{tab:tcs_vs_binary_tcs}
\end{table*}

\begin{table*}[p]
\centering
\small
\renewcommand{\arraystretch}{1.2}
\begin{tabular}{|l|l|l|}
\hline
\textbf{Parameter} & \textbf{CAL} & \textbf{Binary CAL} \\
\hline
Layers & 4 & 2 \\
\hline
Nodes (per layer) & [96, 384, 448, 448] & [72, 88] \\
\hline
Activation (per layer) & [`elu', `softsign', `tanh', `elu'] & [`relu', `swish'] \\
\hline
Dropout (per layer) & [0.6, 0.0, 0.0, 0.6] & [0.0, 0.1] \\
\hline
L1 weight (per layer) & [0.001, 0.0001, 0.0001, 0.0] & [0.0, 0.0] \\
\hline
L2 weight (per layer) & [0.0, 0.0001, 0.0, 0.01] & [0.001, 0.0001] \\
\hline
Epochs & 6000 & 500 \\
\hline
Batch size & 16 & 16 \\
\hline
Optimizer & Nadam & RMSprop \\
\hline
Learning rate & 0.01 & 0.001 \\
\hline
Loss & binary\_crossentropy & binary\_crossentropy \\
\hline
Metric & accuracy & accuracy \\
\hline
Early stopping monitor & val\_loss & val\_loss \\
\hline
Early stopping min\_delta & 1.0$\times 10^{-4}$ & 1.0$\times 10^{-4}$ \\
\hline
Early stopping patience & 15 & 15 \\
\hline
Early stopping mode & auto & auto \\
\hline
\end{tabular}
\caption{Comparison of hyper-tuned and binary-training configurations for the CAL subsystem ANN (values shown per layer where applicable).}
\label{tab:cal_vs_binary_cal}
\end{table*}

\begin{table*}[p]
\centering
\small
\renewcommand{\arraystretch}{1.2}
\begin{tabular}{|l|l|l|}
\hline
\textbf{Parameter} & \textbf{Combiner} & \textbf{Binary Combiner} \\
\hline
Layers & 1 & 1 \\
\hline
Nodes (per layer) & [18] & [18] \\
\hline
Activation (per layer) & [`relu'] & [`relu'] \\
\hline
Dropout (per layer) & [0.0] & [0.0] \\
\hline
L1 weight (per layer) & [0.0] & [0.0] \\
\hline
L2 weight (per layer) & [0.0001] & [0.0001] \\
\hline
Epochs & 6000 & 500 \\
\hline
Batch size & 16 & 16 \\
\hline
Optimizer & Adamax & Adamax \\
\hline
Learning rate & 0.01 & 0.01 \\
\hline
Loss & binary\_crossentropy & binary\_crossentropy \\
\hline
Metric & accuracy & accuracy \\
\hline
Early stopping monitor & val\_loss & val\_loss \\
\hline
Early stopping min\_delta & 1.0$\times 10^{-4}$ & 1.0$\times 10^{-4}$ \\
\hline
Early stopping patience & 15 & 15 \\
\hline
Early stopping mode & auto & auto \\
\hline
\end{tabular}
\caption{Comparison of hyper-tuned and binary-training configurations for the Combiner ANN (values shown per layer where applicable).}
\label{tab:combiner_vs_binary_combiner}
\end{table*}

\bibliography{references}

\end{document}